# Climate Forcing by the Volcanic Eruption of Mount Pinatubo. Revised edition*

David H. Douglass and Robert S. Knox

Department of Physics and Astronomy, University of Rochester
Rochester, NY 14627-0171

We determine the volcano climate sensitivity $\lambda$ and response time $\tau$ for the Mount Pinatubo eruption. This is achieved using observational measurements of the temperature anomalies of the lower troposphere and the aerosol optical density (AOD) in combination with a radiative forcing proxy for AOD. Using standard linear response theory we find $\lambda = 0.18 \pm 0.04$ K/(W/m$^2$), which implies a negative feedback of $-1.0 \pm 0.4$. The intrinsic response time is $\tau = 5.8 \pm 1.0$ months. Both results are contrary to the conventional paradigm that includes long response times and positive feedback. In addition, we analyze the outgoing longwave radiation during the Pinatubo eruption and find that its time dependence follows the forcing much more closely than the temperature, and even has an amplitude equal to that of the AOD proxy. This finding is independent of the response time and feedback results.

---

*Original paper *Douglass and Knox* [2005a] See basis for revision in *Douglass and Knox* [2005a,b], *Robock* [2005] and *Wigley et al.* [2005]. Quantitative results have only minor changes.

# 1. Introduction

One of the primary objectives of climatology is to determine how the various forcings affect the climate of Earth. The essential elements of the climate scenario are:

1) A forcing $\Delta F$ [solar, $CO_2$, $CH_4$, ENSO, volcanoes, etc.] disturbs the climate system;

2) The temperature $T$ of the earth changes by $\Delta T$ with a response time $\tau$;

3) The magnitude of the response is determined by a sensitivity $\lambda$;

4) The forcing $\Delta F$ may have an associated feedback process resulting in a gain $g$ which is a factor in $\lambda$. The gain $g$ can be greater than or less than unity, depending on the sign of the feedback.

The Pinatubo volcano climate event (June 15, 1991) dominated all other forcings during the time of its occurrence. As *Hansen et al.* (1992) said: this dramatic climate event had the potential to "…[exceed] the accumulated forcing due to all anthropogenic greenhouse gases added to the atmosphere since the industrial revolution began"… and should "provide an acid test for global climate models." The temperature of the earth decreased by 0.5 C and the outgoing long wave flux decreased by 2.5 W/m$^2$. We present a new analysis with no adjustable parameters based upon observational data and one theoretical result (volcano forcing) that yields the values of the climate parameter $\lambda = 0.18 \pm 0.06$ K/(W/m$^2$). This implies negative feedback of $-1.6$ ($+0.7, -1.6$), and an intrinsic response time $\tau = 5.8 \pm 1.0$ months. These values are quite different from those that have been assumed or found by previous investigators. We suggest that the reason is that they assumed, either explicitly or implicitly [see, *e. g.*, *Lindzen and*

*Giannitsis,* 1998], that climate relaxation times are long compared to the relevant volcano time scales, an assumption that we do not make and which can be seen to be inconsistent with the observed climate response.

The forcing $\Delta F(t)$ is defined in terms of an equivalent change in net irradiance (in W/m$^2$) referred to the top of the atmosphere [*Shine et al.,* 1995]. This forcing causes a change in the mean temperature of Earth. It is assumed that this formalism applies to $\Delta F(t)$ and $\Delta T(t)$ as global averages. Climate models concentrate on predicting a sensitivity parameter $\lambda$ that connects these quantities,

$$\Delta T(t) = \lambda \Delta F(t), \tag{1}$$

for very slow variations in forcing ("steady state") [*Shine et al.,* 1995]. When the system is not in steady state there is a response time $\tau$ introducing a delay between $\Delta F(t)$ and $\Delta T(t)$. Energy balance models incorporating such a response time have been used for many years [*e. g., North et al.,* 1981], with the dynamics expressed in the form

$$\tau \frac{d\Delta T}{dt} + \Delta T = \lambda \Delta F. \tag{2}$$

*Douglass et al.* (2004a) [6] have shown the connection of Eq. (2) to a two-level atmosphere model in the case of solar forcing and in the presence of explicit (but unspecified) feedback $f$. The result can be expressed as

$$\lambda = g\lambda_0, \tag{3}$$

where $\lambda_0$ is an intrinsic (no-feedback) sensitivity and the gain and feedback are related by

$$g = \frac{1}{1-f} \quad (4)$$

in the usual way [*Peixoto and Oort*, 1992, pp. 26-29]. In work in progress we show that feedbacks associated with short-term atmospheric phenomena are essentially decoupled from surface-to-deep-ocean processes, and that the effect of the latter can be regarded as an effective small correction to the feedback.

## 2. Data

We consider three data sets that clearly show the Pinatubo influence.

***Aerosol optical density* (AOD).** The AOD index (dimensionless) is generally accepted as the proxy for volcano climate forcing. *Hansen et al.* [2002] have shown that

$$\Delta F = A \cdot \text{AOD}, \quad (5)$$

where $A = -21$ W/m$^2$. This value of $A$ is the latest estimate by the Hansen group. There are several prior estimates that have been as high as $-30$ W/m$^2$. We arbitrarily assign an uncertainty of $\pm 0.4$ W/m$^2$. The most recent determination of AOD is by Ammann *et al.* (2003).

***Temperature anomalies*.** We use the global monthly satellite Microwave Sounding Unit lower troposphere temperature (TLT) anomaly data set [*Christy et al.*, 2000] that begins in 1979. *Douglass et al.* [2002, 2004b] have used TLT to determine the solar sensitivity in a multiple regression analysis using solar irradiance, El Niño, and AOD as predictor variables. That analysis produced the relation

$$\Delta(\text{TLT}) = k \cdot \Delta(\text{AOD}), \qquad (6)$$

with $k = -2.9 \pm 0.2$ K/$AOD$ and a delay of 8 months. In addition a modified TLT data set was produced with El Niño and solar effects removed. We designate this as TLTm. Our reported analysis will be based on TLTm with occasional comparisons to analysis based on TLT.

**Long wave radiation (LW).** The outgoing long wave (LW) radiation data are from *Minnis* [1994]. The LW fluxes were determined from irradiance measurements from the Earth Radiation Budget Experiment and are referenced to monthly means from 1985 through 1989. It is noted that the measurements are confined to those made between latitudes 40° N and 40° S, comprising 77% of Earth's surface. We assume that the radiation outside of this band will not seriously change the average flux values reported.

Figure 1(a) shows plots of TLT, TLTm, LW and AOD for the period 1979 to 2003. The LW data cover only 1985 through May 1993, the duration of those measurements. Figure 1(b) shows an expanded plot from 1991 to 1994 emphasizing the period of the Pinatubo volcanic event.

## 3. Analysis

The Pinatubo eruption produced aerosols that reflected solar radiation away from Earth, causing a general change in the energy balance. These events were quantified in the aerosol optical density (AOD) and longwave emission (LW) data sets, respectively. Since $\Delta T$ clearly lags $\Delta F$, it is reasonable to apply the straightforward linear response theory represented by Eq. (2). We use the time dependence of AOD to obtain a solution of this equation, assuming with Hansen that $\Delta F$ is proportional to AOD:

$$\Delta F = -Aq(t), \quad (7)$$

where $q(t)$ is a function that closely fits the AOD data,

$$q(t) = 0.439(t/t_V)\exp(-t/t_V). \quad (8)$$

The time $t$ is in years measured from 1991.42 and $t_V$ is the time that AOD reaches its maximum, in our case 0.63 yr = 7.6 mo. The function $q(t)$ is compared with AOD in Figure 2.

The exact analytic solution of Eq. (2) with the forcing given by Eqs. (7,8) is

$$\Delta T(t) = -0.439\lambda A \cdot \frac{t_V \tau}{(\tau - t_V)^2} \cdot \left\{ \exp(-t/\tau) - \left[\left(\frac{1}{t_V} - \frac{1}{\tau}\right)t + 1\right]\exp(-t/t_V) \right\}, \quad (9)$$

where $\lambda A$ and $\tau$ will be determined by fitting to the $\Delta T$ data set TLTm. By least-squares analysis we obtain a best fit with $\tau = 0.47$ yr and $\lambda A = 3.72$. The fit is shown in Figure 3. The few points near $t = 0$ and many at $t > 6t_V$ lie far outside the predicted value. When we omit these points we find no change in the values of $\tau$ and $\lambda A$. Fitting to TLT, we find $\tau = 0.50$ yr and $\lambda A = 1.97$. Using the value of $A$ determined by *Hansen et al.* [2002], we obtain values of the sensitivity $\lambda$ as derived from the two data sets: $\lambda = 0.18$ (TLTm) and 0.094 (TLT). These incorporate the "direct" value of $A$ and are shown in the first column of Table 2. The close agreement of the relaxation times provides a measure of the accuracy of our dynamical fit.

We have also found the solution of Eq. (2) numerically from the AOD data itself. In the critical region of 0-3 years, the two methods agree with each other closely, having an rms difference of 1.2% of the peak value.

We now consider the LW emission data. It is generally expected to follow the temperature, but we find it to be quite anomalous in the present case. The three data sets, AOD, temperature, and LW emission, are analyzed by the delayed correlation method described by *Douglass et al.* [2004a, 2004b]. We demonstrate the method using the TLT and LW data sets:

(1) LW changes from a background level to a large value and returns.

(2) TLT does the same but is delayed by a time $t_d$.

(3) A plot of TLT against LW will show a "Lissajous loop" whose area is roughly proportional to the time delay. This area would be exactly proportional in the case of sinusoidal functions. For the case of the volcano data sets, the peaks are of different half-widths and there is only a "quarter cycle" of a sinusoid. As a result only the values near the peaks contribute to a loop.

(4) By varying the time lag of one of the variables, LW, one can find the time lag that minimizes the area of the loop. This value of lag determines $t_d$ and also, of interest to us, a linear relation with proportionality constant *s* between TLT and LW. In practice, the linear fit is done by least squares with the slope and the $R^2$ correlation statistic being determined simultaneously. Figure 4(a) shows TLT *vs.* LW, Figure 4(b) shows TLT *vs.* AOD, and Figure 4(c) shows LW *vs.* AOD. The resulting linear relationships are expressed as

$$\Delta(\text{TLT}) = s \cdot \Delta(\text{LW}), \tag{10}$$

$$\Delta(\text{TLT}) = k \cdot \Delta(\text{AOD}), \tag{11}$$

$$\Delta(\text{LW}) = A' \cdot \Delta(\text{AOD}). \tag{12}$$

The values of $s$, $k$, and $A'$ and the associated delays are listed in Table 1. Remarkably, the determined delay between LW and AOD is observed to be 0 and that between LW and TLT is long (6 months), exactly the opposite of what one expects. It is tempting to consider LW as an alternative proxy for the forcing. The delays between TLT/TLTm and LW or AOD are thus expected to be the same. However, there are four different estimates (Table 1). The average and standard deviation are expressed as $t_d = 6.8 \pm 1.5$ months.

The value of $A'$ as determined by regression, Eq. (12), is as remarkable as the zero delay between LW and AOD. Its value is $-21$ W/m$^2$, identical to the theoretical AOD-forcing proxy value $A$. When Eqs. (10), (11), and (12) are compared, we see that $A' = s/k$. This can be called an indirect value of $A$, written $A_{\text{ind}}$. From the slopes of Table 1, we find that $A_{\text{ind}}$ has the value $-16.2$ and $-17.5$, in appropriate units, when evaluated from TLTm and TLT, respectively. These values, being close to the directly determined value of $A'$, reinforce the idea that LW may be a proxy for the forcing. We call $A_{\text{ind}} = s/k$ the "indirect" value of $A'$.

When these results are combined with the two different values of $\lambda A$ determined above, there are four comparable values of the climate sensitivity to consider, as shown in Table 2. The consolidated result is

$$\lambda = (\lambda A)_{\text{data fit}} / A_{\text{regression}} = 0.15 \pm 0.06 \text{ K/(W/m}^2). \tag{13}$$

The numbers refer to the mean and standard deviation of the four values in Table 2, the latter of which is a measure of the systematic error. The value $\lambda = 0.18$ K/(W/m$^2$) derived from the TLTm and direct constant $A$ will be taken as the preferred result of this research. Finally, our estimate of the uncertainty in $A$ (see sec. 2) results in $\lambda = 0.18 \pm 0.04$ K/(W/m$^2$).

## 4. Results and discussion

*Gain and feedback.* The conventional value of sensitivity for global average quantities with radiative forcing and no feedback is the Stefan-Boltzmann value $\lambda_{SB} = 0.30$ K/(W/m$^2$) [*Kiehl*, 1992]. One of the present authors has recently shown [*Knox,* 2004] that the surface-to-atmosphere non-radiative flux makes a correction to the Stefan-Boltzmann result:

$$\lambda_0 = \frac{1}{1-\gamma}\lambda_{SB}, \qquad (14)$$

where $\gamma$ is proportional to the non-radiative flux and has a typical value 0.16. The non-radiative correction is not a feedback effect, so $\lambda_0$ is still properly described as the no-feedback sensitivity, and its value is 20% higher than $\lambda_{SB}$, or 0.36 C/(W/m$^2$).

The gain and feedback of the climate system can now be estimated, using Eqs. (3,4). We have

$$g = \lambda/\lambda_0 = (0.18 \pm 0.04)/0.36 = 0.5 \pm 0.1. \qquad (15)$$

Associated with this gain is a negative feedback,

$$f = -1.0 \pm 0.4. \qquad (16)$$

If the gain and feedback are evaluated without the non-radiative correction, a similar result is obtained, $g = 0.6 \pm 0.1$, with the feedback again always negative.

*Mechanisms.* This work raises the question of the origin of a response time as short as several months. This is just the characteristic time it takes for atmospheric disturbance to propagate over the earth. We conclude that the climate event that begins in the atmosphere remains in the atmosphere and that there is negligible coupling to the deep ocean. In addition, we conclude that there is no "climate left in the pipeline," as discussed below.

Since our analysis yields a gain less than unity, a second issue raised is the origin of the required negative feedback. Negative feedback processes have been proposed involving cirrus clouds [*Lindzen et al.*, 2001]; and *Sassen* [1992] reports that cirrus clouds were produced during the Mt. Pinatubo event. The Lindzen *et al.* process involving clouds yields a negative feedback factor of $f = -1.1$, which is well within the error estimate of the feedback found by us.

Why has no one come to these conclusions before? From the observations, with no analysis at all, one can estimate

$$\Delta T / \Delta F = \Delta T / (-21 \cdot \text{AOD}) = \sim (-0.7) / (-21 \times 0.162) = 0.2 \sim \lambda \qquad (17)$$

and one also sees that the peak of $\Delta T$ occurred about 7 months after the peak in AOD. This is surprisingly close to the values that our detailed analysis yields. We suggest that this solution was rejected because of a widely held belief in a paradigm that assumes that the intrinsic response time is much greater than the volcano event time, mathematically, that $\tau \gg t_V$. This paradigm also includes/induces a belief that positive feedback processes are present. How can the observation be explained within this paradigm? In the limit $\tau \gg t_V$ one sees that the solution Eq. (9) becomes

$$\Delta T(t) \xrightarrow[\tau \gg t_V]{} -0.439 \lambda A \cdot \frac{t_V}{\tau} \cdot \left\{ \exp(-t/\tau) - \left(1 - \frac{t}{\tau}\right) \exp(-t/t_V) \right\}. \quad (18)$$

This result has two relevant features. The first exponential term dominates when $t > t_V$, so the tail of the response drops very slowly, with a characteristic time $\tau$, if $\tau$ is large. This "memory effect" has often been called climate in the pipeline. This is *not* supported by the Pinatubo data. Secondly, the factor $t_V/\tau$ acts as a dynamical factor and reduces the peak value. Now, the above "back of the envelope" calculation becomes

$$\Delta T / \Delta F \sim \lambda \cdot (t_V/\tau). \quad (19)$$

(The final factor is there because in this calculation "$\Delta T$" refers to the peak amplitude, which now contains an effective, smaller $\lambda A$ as seen in Eq. 19.) So if one were to believe that $\tau \sim 3$ to $10\ t_V$, then one would estimate $\lambda \sim 0.5$ to $2$ and infer $g > 1$ and positive feedback. Thus one **"explains"** the observations within the paradigm, but on the basis of a solution of the equations that does not take account of the forcing shape. Note that the proportionality between $\lambda$ and $\tau$ implied by Eq. (19) is guaranteed only in the limit $\tau \gg t_V$. It is not a feature of the exact solution, Eq. (9).

In summary, we have shown that Hansen's hope that the dramatic Pinatubo climate event would provide an "acid test" of climate models has been achieved, although with an unexpected result. The effect of the volcano is to reveal a short atmospheric response time, of the order of several months, leaving no climate in the pipeline, and a negative feedback to its forcing.

# References

Ammann C. M., G. A. Meehl, and W. W. Washington (2003), A monthly and latitudinally varying forcing data set in simulations of the 20th century climate, *Geophys. Res. Lett., 30*, no 12 1657, doi: 10.1029/2003GL016875.

Christy, J. R., R. W. Spencer, and W. D. Braswell (2000), MSU tropospheric temperatures: dataset construction and radiosonde comparisons, *J. Atmos. Oceanic Tech., 17*, 1153-1170. Updates available at http://vortex.nsstc.uah.edu/data/msu/t2lt/.

Douglass D. H and B. D. Clader (2002), Climate Sensitivity of the Earth to Solar Irradiance, *J. Geophys. Res., 29*, doi:10.1029/2002GL015345.

Douglass, D. H., E. G. Blackman, and R. S. Knox, (2004a), Temperature response of Earth to the annual solar irradiance cycle, Phys. Lett. A, 323, 315-322, 2004a; erratum, *Phys. Lett. A*, 325, 175-176, 2004; see revision that incorporates erratum, ArXiv publication http://arXiv.org/abs/astro-ph/0403271

Douglass D. H., B. D. Clader, and R. S. Knox (2004b), Climate sensitivity of Earth to solar irradiance: update. 2004 Solar Radiation and Climate (SORCE) meeting on Decade Variability in the Sun and the Climate, Meredith, New Hampshire, October 27-29, 2004. Available at http://arXiv.org/abs/0411002.

Douglass D.H. and R.S. Knox (2005a), Climate forcing by the volcano eruption of Mount Pinatubo. *Geophys. Res. Lett. 32*, L05710.doi: 10.1029/2004GL022119.

Douglass D.H. and R.S. Knox (2005b), Reply to *Wigley et al.* [2005], *Geophys. Res. Lett.*, in press.

Douglass D.H. and R.S. Knox (2005c), Reply to *Robock* [2005], *Geophys. Res. Lett.*, in press.

Hansen, J., A. Lacis, R. Ruedy, and M. Sato (1992), Potential climate impact of Mount Pinatubo eruption, *Geophys. Res. Letters, 19*, 215-218.

Hansen, J., M. Sato, and R. Ruedy (1997), Radiative forcing and climate response. *J. Geophys. Res. 102*, 6831-6864.

Hansen, J., and 27 others (2002), Climate forcings in Goddard Institute for Space Studies SI2000 simulations, *J. Geophys. Res., 107*(D18), 4347, doi:10.1029/2001JD001143.

Kiehl J. T. (1992), Atmospheric general circulation modeling, in *Climate System Modeling*, edited by K. E. Trenberth, Cambridge Univ. Press, Cambridge, p. 324..

Knox, R. S. (2004), Non-radiative energy flow in elementary climate models, *Phys. Lett. A, 329*, 250-256.

Lindzen, R. S. (2001), Does the earth have an adaptive infrared iris? *Bull. Am. Meteorol. Soc., 82*, 417.

Lindzen, R. S. and C. Giannitsis (1998), On the climatic implications of volcanic cooling, *J. Geophys. Res. 103*, 5929-5941.

Minnis, P. (1994), Radiative Forcing by the 1991 Mt. Pinatubo Eruption, Eighth Conf. on Atmospheric Radiation, Amer. Meteorol. Soc., Boston, MA (Jan 23 - 28, 1994) J9-J11. [tabular data by personal communication]

North G. R., R. F. Cahalan, and J. A. Coakley, Jr. (1981), Energy balance climate models, *Rev. Geophys. and Space Sci. 19*(1), 91-121.

Peixoto, J. P. and A. H. Oort (1992), *Physics of Climate*, American Institute of Physics, New York, pp. 26-29.

Robock, A. [2005], Using the Mount Pinatubo volcanic eruption to determine climate sensitivity: Comments on "Climate forcing by the volcanic eruption of Mount Pinatubo" by David H. Douglass and Robert S. Knox, submitted to Geophys. Res. Lett.

Sassen, K. (1992), Evidence for liquid-phase cirrus cloud formation from volcanic aerosols: climate indications, *Science 257*, 516-519.

Shine K. P., Y. Fouquart, V. Ramaswamy, S. Solomon, and J. Srinivasan (1995), Radiative Forcing. in *Climate Change 1994*, edited by J. T. Houghton *et al.*, Cambridge Univ. Press, Cambridge, pp. 162-204.

Wigley, T. M. L., C. M. Ammann, and B. D. Santer [2005] Using the Mount Pinatubo volcanic eruption to determine climate sensitivity: Comments on "Climate forcing by the volcanic eruption of Mount Pinatubo" by David H. Douglass and Robert S. Knox (Geophys. Res. Lett., 32, L05710, doi:10.1029/2004GL022119, 2005), submitted to Geophys. Res. Lett.

---

D. H. Douglass, Department of Physics and Astronomy, University of Rochester, Rochester, NY 14627-0171 (douglass@pas.rochester.edu)

R. S. Knox, same address (rsk@pas.rochester.edu)

**Table 1**. Values of *s*, *k*, *A´*: the coefficients of the Lissajous linear regressions described in the text. "Delay" is the time from forcing peak to response peak. The last column is the relaxation time as estimated by using Eq. (9); the average and standard deviation are 6.8 ± 1.5 months, consistent with the value found by directly fitting the TLTm curve (5.8 months).

| Coefficient | Value | Delay (mo.) | $\tau$ (mo.) |
|---|---|---|---|
| *s* from TLT *vs* LW | $0.127 \pm 0.033$ K/(W/m$^2$) | 6 | 5.5 |
| *s* from TLTm *vs* LW | $0.182 \pm 0.0486$ K/(W/m$^2$) | 6 | 5.5 |
| *k* from TLT *vs* AOD | $-2.216 \pm 0.298$ K | 8 | 8.4 |
| *k* from TLTm *vs* AOD | $-2.948 \pm 0.42$ K | 7 | 7.6 |
| *A* from LW *vs* AOD | $-21.0 \pm 2.7$ W/m$^2$ | 0 | 0 |

**Table 2.** Values of the climate sensitivity determined from a range of statistical methods, as discussed in the text. The preferred value of $\lambda$ is 0.18 K/(W/m$^2$) because the TLTm data set refers most specifically to volcano data. These units apply to all four cases in the table.

| $\Delta T$ data | $(\lambda A)_{\text{FIT}}/A_{\text{regression}}$ with direct values of $A$ (or $A'$) | $(\lambda A)_{\text{FIT}}/A_{\text{regression}}$ with indirect values of $A'$ |
|---|---|---|
| TLTm | $\lambda = (-3.72)/(-21.0)$ <br> = **0.18** | $\lambda = (-3.72)/(-16.2)$ <br> = **0.23** |
| TLT | $\lambda = (-1.97)/(-21.0)$ <br> = **0.094** | $\lambda = (-1.97)/(-17.5)$ <br> = **0.11** |

**Figure captions**

1. Data sets for temperature (TLT), modified temperature (TLTm), aerosol optical density (AOD), and outgoing long wave radiation (LW). the modified data set has the El Niño and solar signals removed (see text). (a) Complete sets, (b) expanded view showing the subsets used in the Pinatubo analysis.

2. Volcano AOD function (detail of AOD from Fig. 1) and the analytic fit $0.162q(t)$ (text, Eq. 8).

3. Fit of the analytic solution $\Delta T(t)$, Eq. 9, to the temperature data set TLTm.

4. Lissajous patterns used to evaluate delays and amplitudes of pairs of data sets. (a) TLT and LW; (b) TLT and AOD; (c) LW and AOD.

DOUGLASS-KNOX FIG. 1

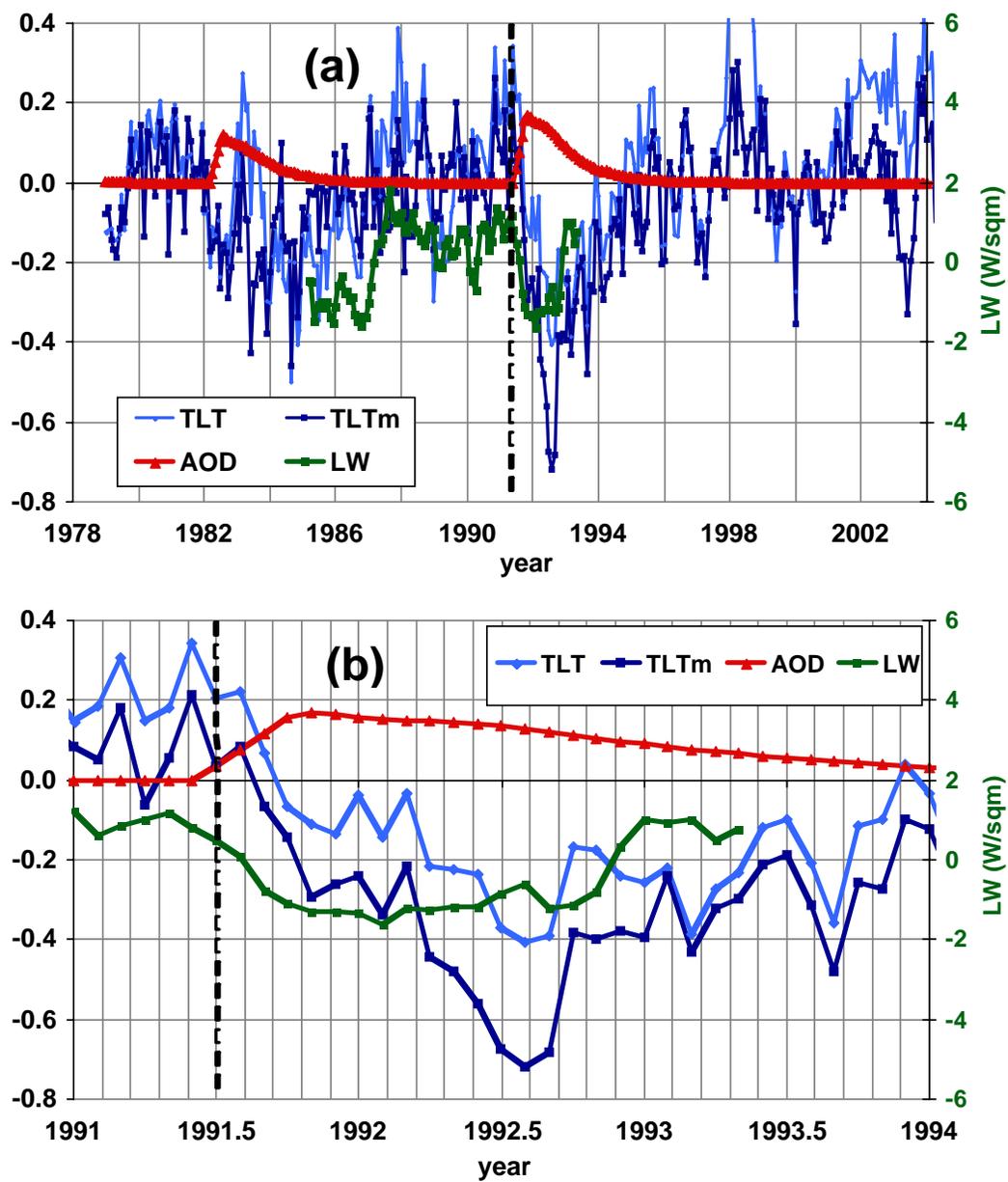

DOUGLASS-KNOX   FIG. 2

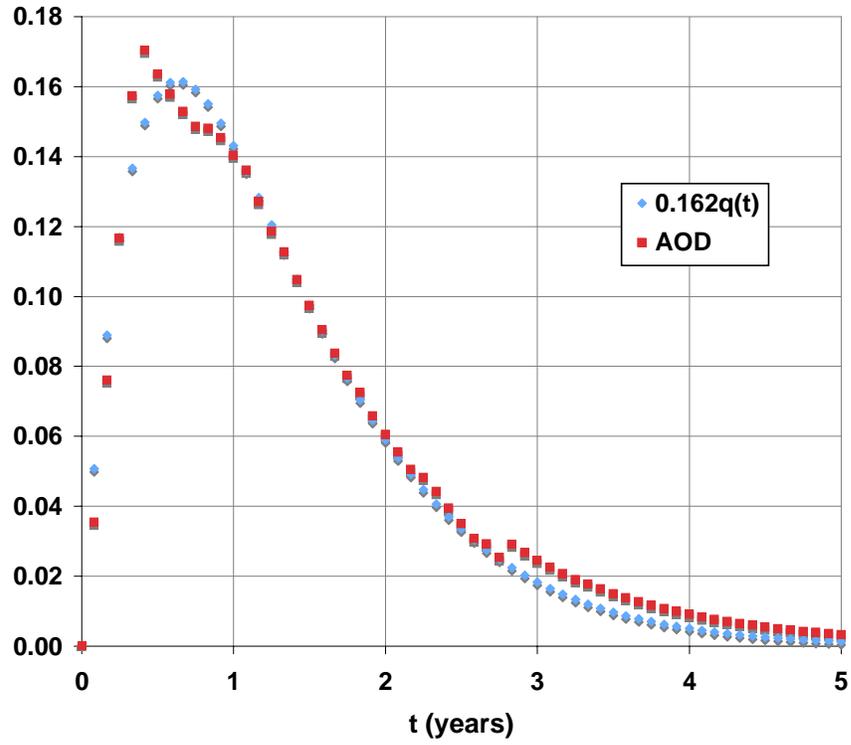

DOUGLASS-KNOX FIG. 3

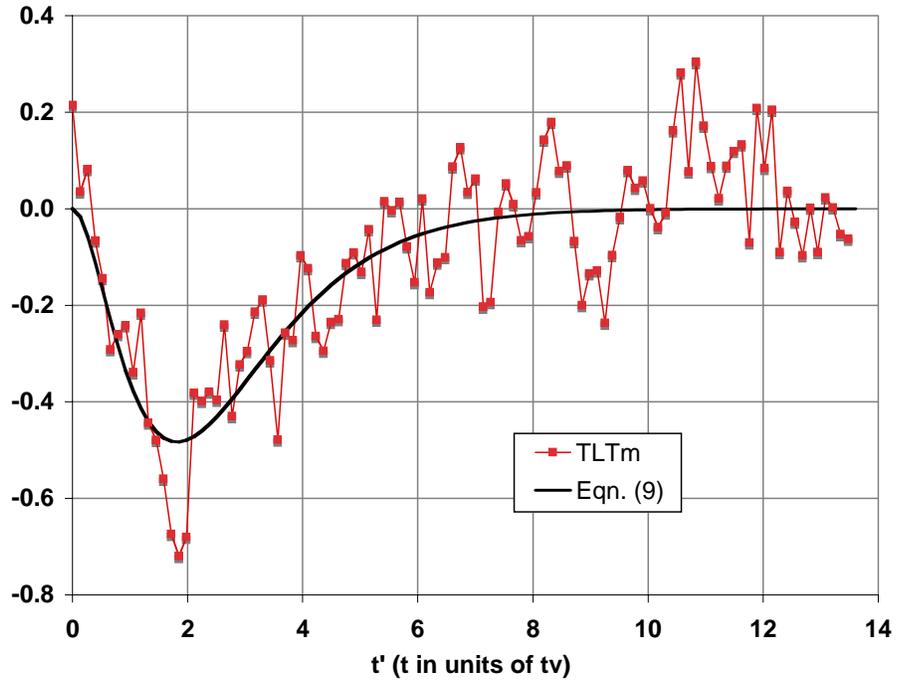

DOUGLASS-KNOX FIG. 4

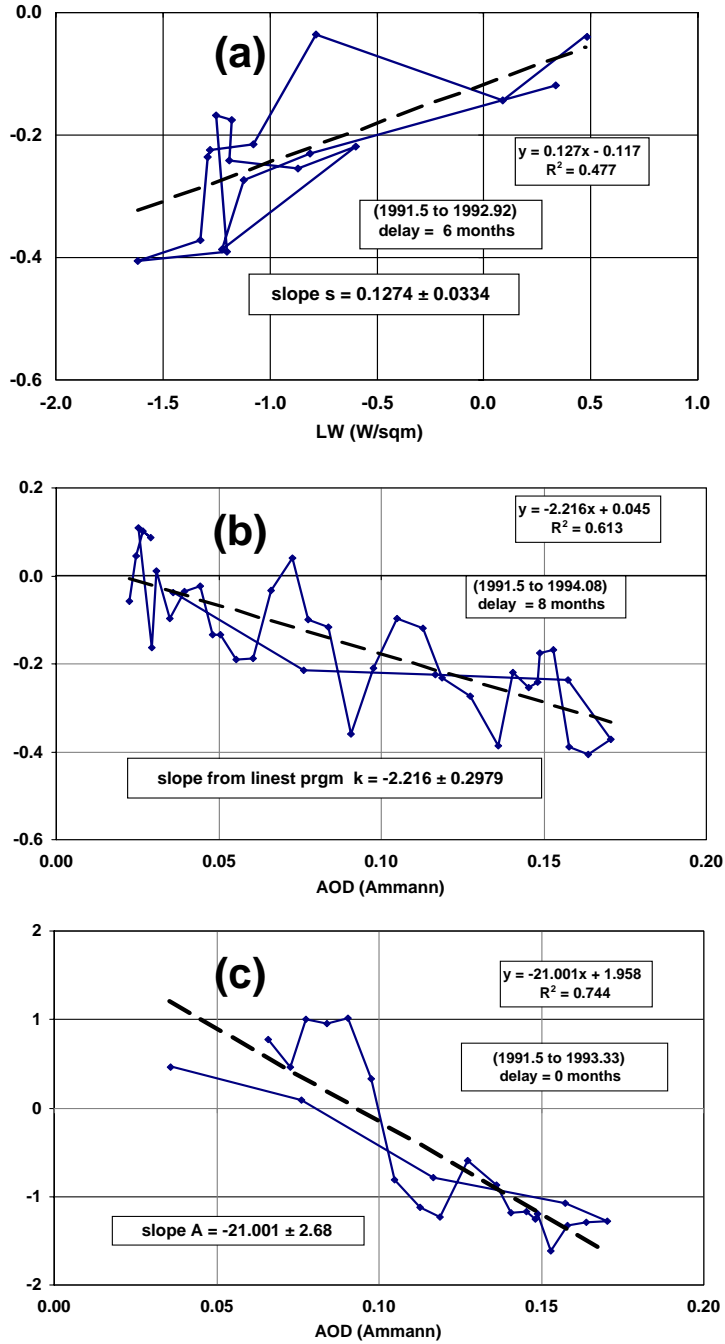